\documentclass[prd,showpacs,showkeys,floatfix,nofootinbib,%%eqsecnum,
               preprint,12pt,tightenlines,fleqn]{revtex4-1}
\usepackage{amssymb,epsf}
\usepackage{latexsym,,amssymb}
\usepackage{amsmath,mathrsfs}
\usepackage{graphics,epsfig}
\usepackage[english]{babel}
\newcommand{\beq}{\begin{equation}}
\newcommand{\eeq}{\end{equation}}
\newcommand{\beqa}{\begin{eqnarray}}
\newcommand{\eeqa}{\end{eqnarray}}
\newcommand{\bsubeqs}{\begin{subequations}}
\newcommand{\esubeqs}{\end{subequations}}
\begin{document}
\noindent  Phys. Rev. D 90, 024007 (2014)
\hfill arXiv:1402.7048%%\;(\version)
\newline\vspace*{2mm}
%
%\noindent  arXiv:1402.7048
%\hfill  KA--TP--08--2014\;(\version)\newline\vspace*{2mm}
%
\title{\vspace*{2mm}Skyrmion spacetime defect\\[2mm]\vspace*{2mm}}
\author{F.R.~Klinkhamer}
\email{frans.klinkhamer@kit.edu}
\affiliation{Institute for
Theoretical Physics, Karlsruhe Institute of
Technology (KIT), 76128 Karlsruhe, Germany\\}

\begin{abstract}%
\noindent
\vspace*{-4mm}\newline
A finite-energy static classical solution is obtained for
standard Einstein gravity coupled to an $SO(3)\times SO(3)$ chiral model
of scalars [a Skyrme model].  
This nonsingular localized solution has nontrivial topology
for both the spacetime manifold and the $SO(3)$ matter fields.
The solution corresponds to a single
spacetime defect embedded in flat Minkowski spacetime.
\end{abstract}


\maketitle

\section{Introduction}\label{sec:Introduction}

The classical spacetime emerging from a quantum-spacetime phase
may very well have nontrivial structure at small length
scales~\cite{Wheeler1957,Wheeler1968,Visser1995}.
This nontrivial structure affects, in particular, the propagation of
electromagnetic waves~\cite{BernadotteKlinkhamer2006}.
The question remains as to what the small-scale
structure embedded in a flat spacetime really looks like (here, 
we do not consider intra-universe wormhole solutions which
connect \emph{two} asymptotically flat spaces~\cite{Visser1995}).

Narrowing down the question, is it
possible at all to have nonsingular localized finite-energy solutions of
the standard Einstein equations with
nontrivial topology on small length scales and
flat spacetime asymptotically?
For one particular topology studied in Ref.~\cite{BernadotteKlinkhamer2006},
it has been suggested~\cite{Schwarz2010}
to consider an $SO(3)$ Skyrme model
coupled to gravity~\cite{Skyrme1961,MantonSutcliffe2004,LuckockMoss1986,%
Glendenning-etal1988,DrozHeuslerStraumann1991,BizonChmaj1992,%
HeuslerStraumannZhou1993,KleihausKunzSood1995}.

For this specific theory and topology,
two nonsingular defect solutions have been
found recently, a vacuum solution and a nonvacuum
solution~\cite{KlinkhamerRahmede2013}.
(Corresponding nonsingular black-hole solutions were presented
in Refs.~\cite{Klinkhamer2013-MPLA,Klinkhamer2013-APPB}.)
Both defect solutions of Ref.~\cite{KlinkhamerRahmede2013}
are localized, but the one with a nonvanishing scalar $SO(3)$
field has infinite energy and trivial scalar-field-configuration
topology (i.e., zero winding number or ``baryon'' charge).
It has been conjectured~\cite{KlinkhamerRahmede2013} that
this particular nonvacuum solution is unstable
and, by emitting outgoing waves of scalars,
decays to a finite-energy Skyrmion-like defect solution
(with unit winding number or ``baryon'' charge).
Such a nonsingular defect solution with unit winding number
is constructed in the present article.

The construction of this finite-energy Skyrmion spacetime  defect solution
turns out to be quite subtle.
Essential is a perfect control of the fields
at the defect core, which, in turn, requires the
use of appropriate coordinates and fields.
These preliminaries are discussed in Sec.~\ref{sec:Theory}.
The numerical solution is then given in Sec.~\ref{sec:Numerical-solution}.  
Concluding remarks are presented in Sec.~\ref{sec:Discussion}. 
Two technical issues are dealt with in appendices
\ref{app:Alternative-Ansaetze} and \ref{app:Reduced-expressions}.

%%\newpage%%tmp
\section{Theory}\label{sec:Theory}

The setup of the theory has been described elsewhere in
detail~\cite{KlinkhamerRahmede2013,Klinkhamer2013-MPLA,Klinkhamer2013-APPB,Klinkhamer2013-review},
but, for completeness, we recall the main steps.

\subsection{Manifold}
\label{subsec:Manifold}

The four-dimensional spacetime manifold considered in this article has
the topology
\bsubeqs\label{eq:M4-M3-topology}
\beq\label{eq:M4}
M_4 = \mathbb{R} \times M_3\,.
\eeq
The three-space $M_3$ carries the nontrivial topology
and is, in fact,  a noncompact, orientable, nonsimply-connected 
manifold without boundary. Up to a point, $M_3$ is homeomorphic
to the three-dimensional real-projective space,
\beq\label{eq:M3}
M_3 \simeq
\mathbb{R}P^3 - \{\text{point}\}\,.
\eeq
\esubeqs
Adding the ``point at infinity'' gives the compact
space $\overline{M_3} \simeq \mathbb{R}P^3$.

\begin{figure*}[t] %%[t] or [b] or [p]
\includegraphics[width=0.5\textwidth]
{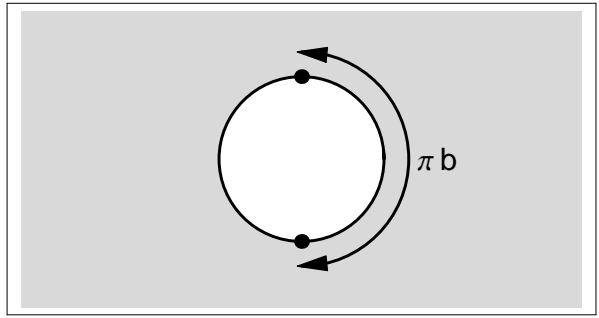}  
\caption{Three-space $M_3$ obtained by surgery on
three-dimensional Euclidean space $E_3$:
the interior of the ball with radius $b$ is removed
and antipodal points on the boundary of the ball are
identified (as indicated by the dots). 
The ``long distance'' between antipodal points is $\pi\, b$ in $E_3$.}
\label{fig:defect}
\end{figure*}

For the direct construction of $M_3$, we perform local
surgery on  the three-dimensional Euclidean space
$E_3=\big(\mathbb{R}^3,\, \delta_{mn}\big)$.
We use the standard Cartesian and spherical
coordinates on $\mathbb{R}^3$,
\beq\label{eq:Cartesian-spherical-coord}
\vec{x}
\equiv |\vec{x}|\, \widehat{x}
= (x^1,\,  x^2,\, x^3)
= (r \sin\theta  \cos\phi,\,r \sin\theta  \sin\phi,\, r \cos\theta )\,,
\eeq
with $x^m \in (-\infty,\,+\infty)$,
$r \geq 0$, $\theta \in [0,\,\pi]$, and $\phi \in [0,\,2\pi)$.
Now, we obtain  $M_3$ from $\mathbb{R}^3$ by removing
the interior of the ball $B_b$ with radius $b$ and
identifying antipodal points on the boundary $S_b \equiv \partial B_b$.
Denoting point reflection by $P(\vec{x})=-\vec{x}$,
the three-space $M_3$ is given by
\beq\label{eq:M3-definition}
M_3 =
\big\{  \vec{x}\in \mathbb{R}^3\,:\; \big(|\vec{x}| \geq b >0\big)
\wedge
\big(P(\vec{x})\cong \vec{x} \;\;\text{for}\;\;  |\vec{x}|=b\big)
\big\}\,,
\eeq
where $\cong$ stands for pointwise identification (Fig.~\ref{fig:defect}).
A minimal noncontractible loop (with length $\pi b$ in $E_3$)
corresponds to half of a great circle on the $r=b$ sphere in $E_3$,
taken between antipodal points which are to be identified.

%%\newpage%%tmp
The single set of coordinates \eqref{eq:Cartesian-spherical-coord}
does not suffice for an appropriate description of $M_3$.
The reason is that two different values of these coordinates may
correspond to one point  of $M_3$ as defined in \eqref{eq:M3-definition}.
An example is provided by the two sets of coordinates
$\vec{x}=(0,\,b,\,0)$ and $\vec{x}=(0,\,-b,\,0)$,
which describe the same point of $M_3$ (see Fig.~\ref{fig:defect}).

A particular covering of $M_3$ uses three charts of coordinates,
labeled by $n=1,2,3$. The basic idea~\cite{Schwarz2010} is
that the coordinates resemble spherical coordinates and that
each coordinate chart surrounds one of the three Cartesian coordinate axes
but does not intersect the other two axes. These coordinates are denoted
\beq\label{eq:XnYnZn}
(X_n,\,  Y_n,\, Z_n)\,, 
\eeq
for  $n=1,\,2,\,3$. In each chart, 
there is one polar-type angular coordinate of finite range,
one azimuthal-type angular coordinate of finite range, and
one radial-type coordinate with infinite range.
Specifically, the coordinates have the following ranges:
\bsubeqs\label{eq:XnYnZn-ranges}
\beqa\label{eq:X1Y1Z1-ranges}
X_{1} \in (-\infty,\,\infty) \,,\quad
Y_{1} \in (0,\,\pi)\,,\quad
Z_{1} \in (0,\,\pi)\,,
\eeqa
\beqa\label{eq:X2Y2Z2-ranges}
X_{2} \in (0,\,\pi)\,,\quad
Y_{2} \in (-\infty,\,\infty)\,,\quad
Z_{2} \in (0,\,\pi)\,,
\eeqa
\beqa\label{eq:X3Y3Z3-ranges}
X_{3} \in  (0,\,\pi)\,,\quad
Y_{3} \in (0,\,\pi)\,,\quad
Z_{3} \in  (-\infty,\,\infty)\,.
\eeqa
\esubeqs
The different charts overlap in certain regions: see Appendix~B of
Ref.~\cite{Klinkhamer2013-APPB} for further details.

%%\newpage%%tmp
\subsection{Fields and interactions}
\label{subsec:Fields-and-interactions}

The spacetime manifold \eqref{eq:M4-M3-topology} of the previous section 
is now supplemented with a metric, $g_{\mu\nu}(X)$, whose dynamics is
governed by the standard Einstein--Hilbert action.
In addition, we consider a scalar field $\Omega(X)\in SO(3)$, 
with self-interactions determined by a quartic Skyrme term in the matter
action~\cite{Skyrme1961,MantonSutcliffe2004}.

The combined action of the pure-gravity sector
 and the matter sector is given by ($c=\hbar=1$)
\begin{equation}\label{eq:action}
 S[g,\,\Omega] =\int_{M_4} d^4X\,\sqrt{-g}\,
\Bigg[
\frac{1}{16\pi G_N}\:R
+\frac{f^2}{4}\:\text{tr}\big(\omega_\mu\,\omega^\mu\big)
+\frac{1}{16 e^2}\: \text{tr}\Big(\left[\omega_\mu,\,\omega_\nu\right]
\left[\omega^\mu,\,\omega^\nu\right]\Big)\Bigg]\,,
\end{equation}
in terms of the Ricci curvature scalar $R$ and the
effective vector field
$\omega_\mu \equiv \Omega^{-1}\,\partial_\mu\,\Omega$.
The kinetic term
of the scalar field involves the mass scale $f$.
The Skyrme term corresponds to the trace of the square of the
commutator $\left[\omega_\mu,\,\omega_\nu\right]$ and has the
dimensionless coupling constant $1/e^2$.

The $SO(3)\times SO(3)$ global symmetry of the matter sector
is realized on the scalar field by the following transformation
with constant parameters $S_L,\, S_R \in SO(3)$:
\beq
 \Omega(X) \to S_L \cdot \Omega(X) \cdot S_R^{-1}\,,
\eeq
where the central dot denotes matrix multiplication.
The generic argument $X$ of the fields
$g_{\mu\nu}(X)$ and $\Omega(X)$ and
the measure $d^4 X$ in the integral \eqref{eq:action}
correspond to one of the three different
coordinate charts needed to cover $M_4$.

%%\newpage%%tmp
\subsection{Ans\"{a}tze}
\label{subsec:Ansaetze}

A spherically symmetric \textit{Ansatz}
for the metric is given by the following line element:  
\bsubeqs\label{eq:metric-Ansatz-W-definition} 
\beqa\label{eq:metric-Ansatz}
 ds^2\;\Big|_\text{chart-2} &=&-
\big[\widetilde{\mu}(W)\big]^2\, dT^2
 +\big(1-b^2/W\big)\,\big[\widetilde{\sigma}(W)\big]^2\,(dY_2)^2
\nonumber\\&&
+W \Big[(dZ_2)^2+\sin^2 Z_2\, (dX_2)^2 \Big]\,,
\\[2mm]
\label{eq:W-definition}
W\;\Big|_\text{chart-2} &\equiv& b^2+(Y_2)^2\,,
\eeqa
\esubeqs
and similarly for the chart-1 and chart-3
domains~\cite{Klinkhamer2013-APPB,Klinkhamer2013-review}.
The focus on chart-2 coordinates in \eqref{eq:metric-Ansatz-W-definition}
is purely for cosmetic reasons. Observe that this \textit{Ansatz},
for finite values of $\widetilde{\sigma}(b^2)$, sets the ``short distance''
between the antipodal points of Fig.~\ref{fig:defect} to  zero,
while keeping the ``long distance'' equal to $\pi\, b$.

The scalar field is described by a Skyrmion-type
\textit{Ansatz}~\cite{Schwarz2010,Skyrme1961,MantonSutcliffe2004},
\bsubeqs\label{eq:hedgehog-Ansatz}
\begin{eqnarray}\label{eq:hedgehog-Ansatz-Omega}   
\Omega &=&
\cos\big[\widetilde{F}\left(r^2\right)\big]\;\openone_{3}
-\sin\big[\widetilde{F}\left(r^2\right)\big]\;
\widehat{x}\cdot \vec{S}
%%\nonumber\\&&
+\big(1-\cos\big[\widetilde{F}\left(r^2\right)\big]\big)\;
\widehat{x} \otimes \widehat{x}\,,
\\[2mm]\label{eq:hedgehog-Ansatz-bcs}
\widetilde{F}(b^2) &=& \pi\,,\quad \widetilde{F}(\infty) = 0\,,
\\[2mm]
S_1 &\equiv&  \left(
                \begin{array}{ccc}
                  0 & 0 & 0 \\
                  0 & 0 & 1 \\
                  0 & -1 & 0 \\
                \end{array}
              \right)\,,
\quad
S_2 \equiv  \left(
                \begin{array}{ccc}
                  0 & 0 & -1 \\
                  0 & 0 & 0 \\
                  1 & 0 & 0 \\
                \end{array}
              \right)\,,
\quad
S_3 \equiv  \left(
                \begin{array}{ccc}
                  0 & 1 & 0 \\
                  -1 & 0 & 0 \\
                  0 & 0 & 0 \\
                \end{array}
              \right)\,,
\end{eqnarray}
\esubeqs
with a hedgehog term proportional to $\sin\widetilde{F}$
in \eqref{eq:hedgehog-Ansatz-Omega}
and further terms involving the $3 \times 3$ unit
matrix $\openone_{3}$ and another matrix which reads
in components
$(\widehat{x}\otimes \widehat{x})^{ab}= \widehat{x}^a\, \widehat{x}^b$.

The boundary condition \eqref{eq:hedgehog-Ansatz-bcs} at 
$r \equiv |\vec{x}|=b$
sets the hedgehog term $\widehat{x}\cdot \vec{S}$ in
\eqref{eq:hedgehog-Ansatz-Omega}  to zero and
makes it  possible to employ the single coordinate
chart \eqref{eq:Cartesian-spherical-coord}.
Specifically, there is the following equality:  
\beq
\Omega\big(r,\,\widehat{x}\big)\,\big|_{r=b} =
-\;\openone_{3} +2\; \widehat{x} \otimes \widehat{x}=
\Omega\big(r,\, -\widehat{x}\big)\,\big|_{r=b}\,,
\eeq
which gives the same scalar field $\Omega$ for
antipodal points on the $r=b$ sphere in $\mathbb{R}^3$,
allowing these antipodal points to be identified
in order to obtain $M_{3}$.\footnote{\label{ftn:su2skymion}An
$SU(2)$ Skyrmion \textit{Ansatz}
$U(r,\,\widehat{x})$ $=$ $\cos[\widetilde{G}(r^2)]\;\openone_{2}
+i\,\sin[\widetilde{G}(r^2)]\;\widehat{x}\cdot \vec{\tau}$
with boundary conditions
$\widetilde{G}(b^2) = \pi$ and $\widetilde{G}(\infty) = 0$
would also have the property $U(b,\,\widehat{x})=U(b,\,-\widehat{x})$,
but its energy would be approximately twice that of the $SO(3)$ Skyrmion.}
In order to match the coordinates
used for the metric \eqref{eq:metric-Ansatz-W-definition}, we make
the identification $r^2 = b^2+(Y_2)^2$, and
the explicit relations between $\widehat{x}$ and $(X_2,\, Z_2)$
can be found in  Refs.~\cite{Klinkhamer2013-APPB,Klinkhamer2013-review}.

The \textit{Ansatz} \eqref{eq:hedgehog-Ansatz}
corresponds to a topologically nontrivial scalar field configuration,
a Skyrmion-like configuration with a unit winding number.  
Having an integer winding number for the scalar field configuration
$\Omega(X) \in SO(3)$
occurs because of the homeomorphism $M_3 + \{\text{point}\}\simeq SO(3)$.
In fact, the winding number or topological degree of the compactified
map $\Omega: \overline{M}_3 \to SO(3)$ turns out to be given
by~\cite{MantonSutcliffe2004,Schwarz2010}
\beq\label{eq:deg-Omega}
\text{deg}[\Omega]=
-\frac{2}{\pi}\;\int_{\pi}^{0} d\widetilde{F}\, \sin^2(\widetilde{F}/2)=1\,,
\eeq
where the endpoints of the integral on the right-hand side
correspond to boundary conditions \eqref{eq:hedgehog-Ansatz-bcs}.

Alternative \textit{Ans\"{a}tze} based on Painlev\'{e}--Gullstrand-type
coordinates are presented in Appendix~\ref{app:Alternative-Ansaetze}.

%%\newpage%%tmp
\subsection{Reduced field equations}
\label{subsec:Reduced-field-equations}

At this moment, we introduce
the following dimensionless model parameters and dimensionless  variables:
\bsubeqs\label{eq:dimensionless}
\begin{eqnarray}\label{eq:dimensionless-eta}  
\widetilde{\eta}&\equiv&8\pi\eta \equiv 8\pi\, G_N\, f^2\,,\\[2mm]
w  &\equiv& (e\,f)^2\;W = (y_0)^2+y^2\,,\\[2mm]
y  &\equiv& e\,f\;Y_2\,,\\[2mm]
y_0 &\equiv& e\,f\;b\,.
\end{eqnarray}
\esubeqs
Inserting the \textit{Ans\"{a}tze} of Sec.~\ref{subsec:Ansaetze}
into the Einstein and matter field equations
from the action \eqref{eq:action} gives the corresponding reduced
expressions~\cite{Glendenning-etal1988,Schwarz2010,KlinkhamerRahmede2013}.
From these equations written in terms of the
dimensionless variables \eqref{eq:dimensionless}, we obtain the following
three ordinary differential equations (ODEs):
\bsubeqs\label{eq:final-ODEs}
\begin{eqnarray}
\label{eq:final-ODEs-sigmatilde}
\widetilde{\sigma}'(w)
&=& \widetilde{\sigma}(w)\;
\Big(-\Big[ w - 2 \widetilde{\eta}\,
\Big(1 + 2 w -  \cos[\widetilde{F}(w)]\Big)\,\sin^2[\widetilde{F}(w)/2]\Big]\,
\widetilde{\sigma}(w)^2
\nonumber\\&&
+ w \,\Big[1 + 2 \widetilde{\eta}\,\Big(2\,w^3 (1 +  y_{0}^2)
- 2 y_{0}^2 \,(1 +  y_{0}^2) (y_{0}^4 - 1)
%\nonumber\\&&
+ 2 w\, y_{0}^2 \,(3 y_{0}^2 + 3 y_{0}^4 - 1)
\nonumber\\&&
- w^2 \,(6 y_{0}^2 + 6 y_{0}^4 - 1)
- 2 w \cos[\widetilde{F}(w)]\Big)\,\widetilde{F}'(w)^2\Big]\Big)\Big/
\Big(4 w^2\Big)\,,
%\\[2mm]
\end{eqnarray}
\begin{eqnarray}
\label{eq:final-ODEs-mutilde}
\widetilde{\mu}'(w)
&=&
\widetilde{\mu}(w)\;
\Big(\Big[ w - 2 \widetilde{\eta}\,
\Big(1 + 2 w -  \cos[\widetilde{F}(w)]\Big)\,\sin^2[\widetilde{F}(w)/2]\Big]\,
\widetilde{\sigma}(w)^2
\nonumber\\&&
+ w\,\Big[-1 + 2 w\,\widetilde{\eta}\,\Big(2 + w - 2 \cos[\widetilde{F}(w)]\Big)\,
\widetilde{F}'(w)^2\Big]\Big)\Big/
\Big(4 w^2\Big)
\,,
%\\[2mm]
\end{eqnarray}
\begin{eqnarray}
\widetilde{F}''(w)
&=&
\Big(
\Big[ 3\, \Big(1 +  w - \cos[\widetilde{F}(w)]\Big) \sin[\widetilde{F}(w)]
- \Big(2 + w - 2 \cos[\widetilde{F}(w)]\Big)
\nonumber\\&&
\times  \Big(3 w - 2 \widetilde{\eta} (3 + 6 w - 3 \cos[\widetilde{F}(w)])
\sin^2[\widetilde{F}(w)/2]\Big)\,\widetilde{F}'(w)\Big]\,\widetilde{\sigma}(w)^2
\nonumber\\&&
- 6 \,w^2\,
\Big[1 +  \sin[\widetilde{F}(w)] \widetilde{F}'(w)\Big]\,
\widetilde{F}'(w)\Big)\Big/
%\nonumber\\&&
\Big(2 w^2 \Big[6 + 3 w - 6 \cos[\widetilde{F}(w)]\Big]\Big)
\,,
\end{eqnarray}
\esubeqs
where the prime stands for differentiation with respect to $w$.

The ODEs \eqref{eq:final-ODEs} are to be solved with the following
boundary conditions on the functions $\widetilde{F}(w)$,
$\widetilde{\sigma}(w)$, and  $\widetilde{\mu}(w)$:
\bsubeqs\label{eq:BCS-MATH}
\begin{eqnarray}\label{eq:BCS-MATH-F}
\widetilde{F}(y_{0}^2)&=&  \pi\,,
\quad
 \widetilde{F}(\infty) = 0\,,
\\[2mm]
\label{eq:BCS-MATH-sigmatilde-mutilde}
\widetilde{\sigma}(\infty)&=& 1\,,
\quad
\widetilde{\mu}(\infty)= 1\,.
\end{eqnarray}
\esubeqs
\noindent The three boundary conditions at infinity   
provide for asymptotic flatness and the boundary condition
$\widetilde{F}=  \pi$ at the core gives a topologically nontrivial
scalar field configuration, as discussed in Sec.~\ref{subsec:Ansaetze}.

Different from the ODEs used in Ref.~\cite{KlinkhamerRahmede2013},
the ODEs in the form presented in \eqref{eq:final-ODEs}
have no singularities at the defect core $w=y_0^2 >0$.
As such, they are directly amenable to numerical analysis.

%%\newpage%%tmp
\section{Numerical solution}
\label{sec:Numerical-solution}

\begin{figure*}[p]
\vspace*{-4mm}
\includegraphics[width=0.825\textwidth]%
{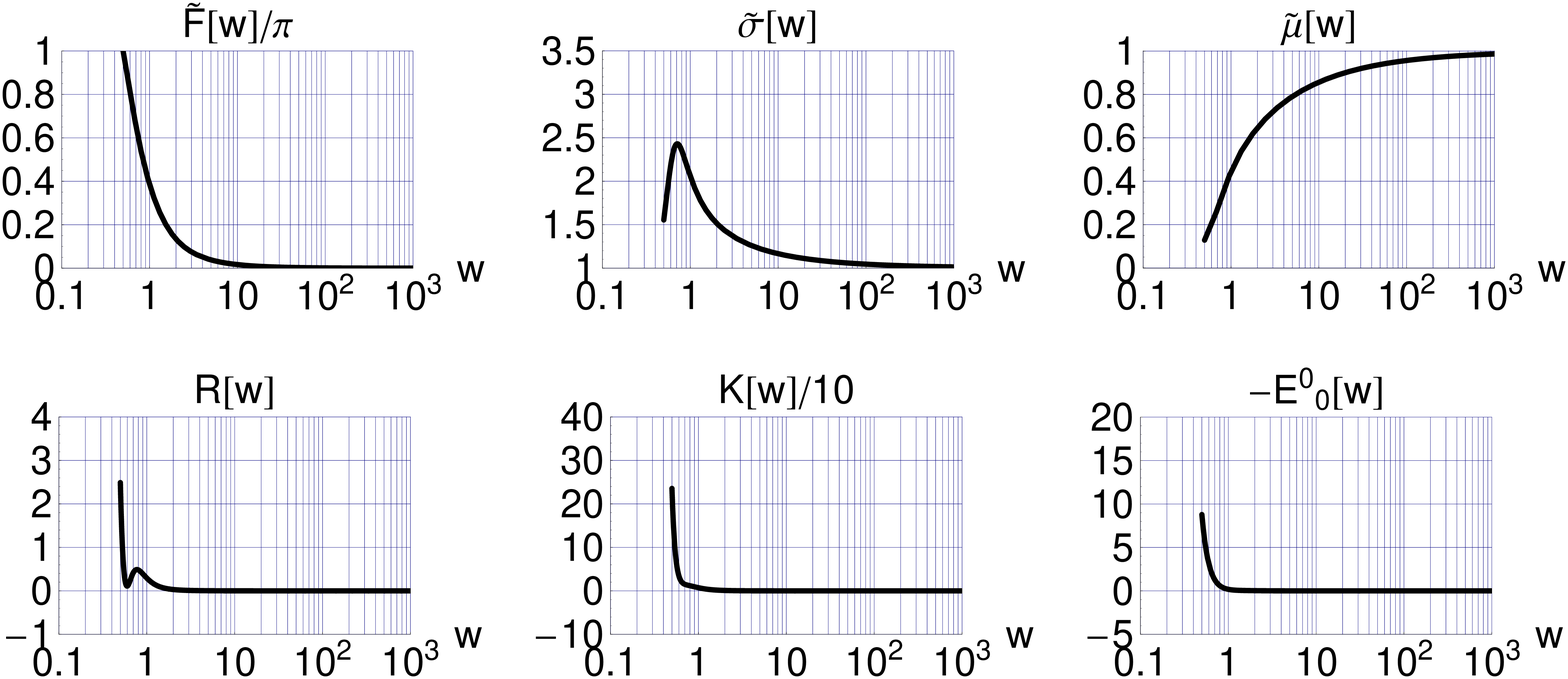}
\vspace*{-4mm}
\caption{\label{fig:functions-bmid}
(Top row) Skyrme function $\widetilde{F}(w)$
and metric functions $\widetilde{\sigma}(w)$
and $\widetilde{\mu}(w)$ of the numerical solution
of the reduced field equations \eqref{eq:final-ODEs}.
The model parameter is $\widetilde{\eta}=1/20$ and
the solution parameter $y_0=1/\sqrt{2}$.
The boundary conditions at the defect core $w=y_0^2=1/2$ are:
$\widetilde{F}=\pi$,
$\widetilde{F}^\prime=-6.3797$,
$\widetilde{\sigma}=1.554$,
and  $\widetilde{\mu}=0.128207$.
(Bottom row) Corresponding dimensionless Ricci scalar $R$,
dimensionless Kretschmann scalar $K$, and the negative of the 
00 component of the dimensionless Einstein tensor $E^{\mu}_{\;\;\nu}$
$\equiv$ $R^{\mu}_{\;\;\nu} -(1/2)\,R\,\delta^{\mu}_{\;\;\nu}$;
see Appendix~\ref{app:Reduced-expressions} for details.
}
\vspace*{1mm}
\includegraphics[width=0.825\textwidth]
{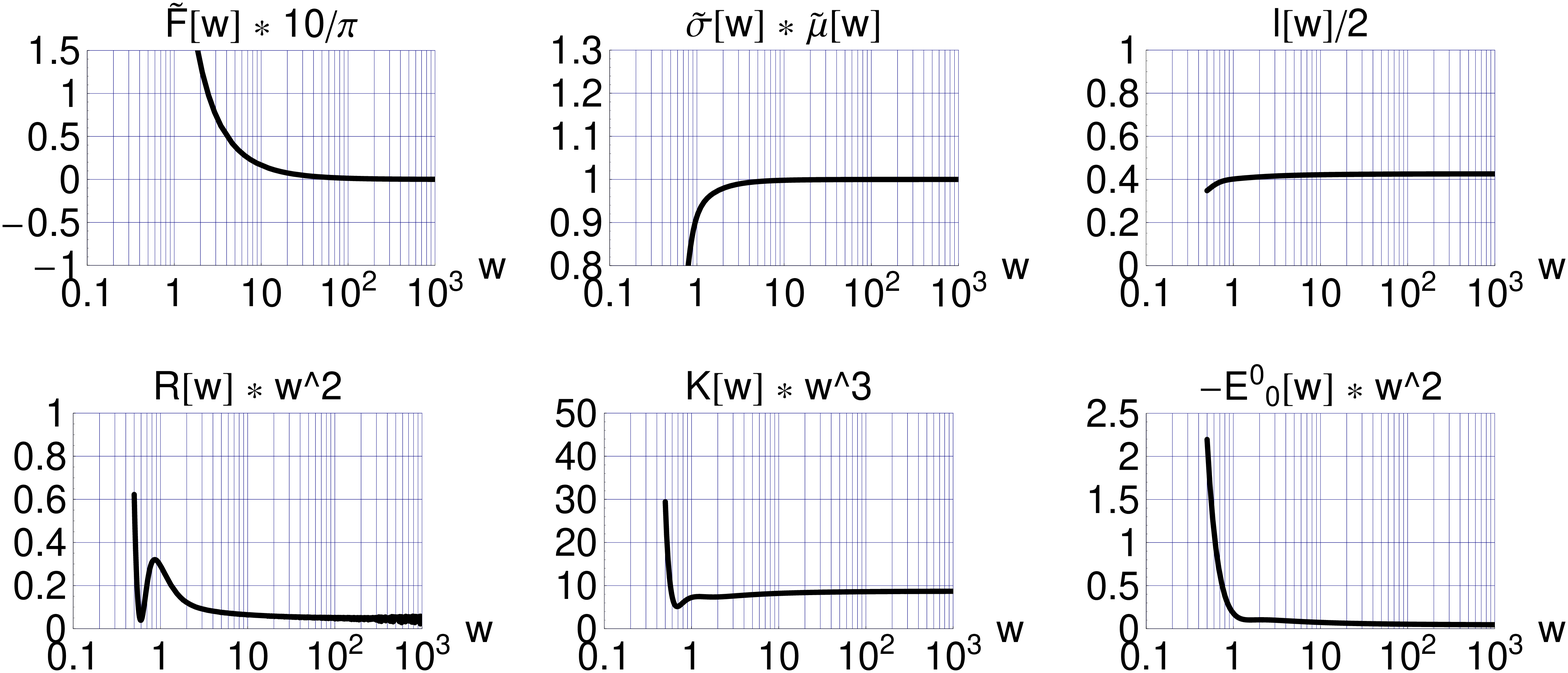}
\vspace*{-4mm}
\caption{\label{fig:asymptotics-bmid}
Asymptotic behavior of the functions from Fig.~\ref{fig:functions-bmid},
with definition $l(w) \equiv  [1 - \widetilde{\mu}(w)^2] \,\sqrt{w}$.
}
\vspace*{1mm}
\includegraphics[width=0.825\textwidth]
{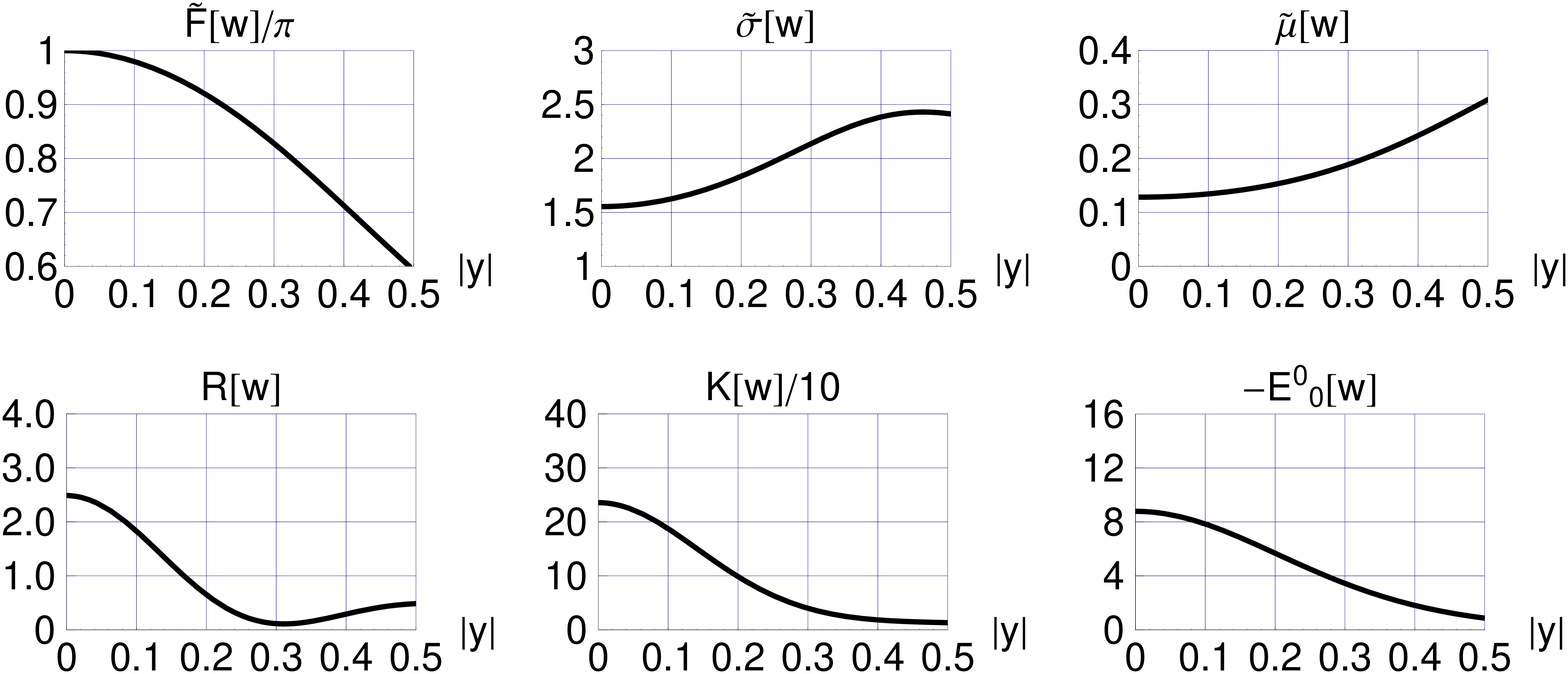}
\vspace*{-4mm}
\caption{\label{fig:core-bmid}
Core behavior of the functions from Fig.~\ref{fig:functions-bmid},
plotted with respect to the dimensionless quasi-radial coordinate
$y \in \mathbb{R}$ of the chart-2 domain. The coordinate $w$
is given by $w= y_0^2+y^2$.}
\vspace*{0mm}
\end{figure*}

In order to obtain the numerical solution corresponding to the
\textit{Ans\"{a}tze} from Sec.~\ref{subsec:Ansaetze},
we have adopted the following procedure.
The boundary conditions on the three functions
$\widetilde{F}(w)$, $\widetilde{\sigma}(w)$, and  $\widetilde{\mu}(w)$
are set at the defect core $w=y_0^2$.  Specifically, we take
\bsubeqs\label{eq:BCS-NUM}
\begin{eqnarray}\label{eq:BCS-NUM-F}
\widetilde{F}(y_{0}^2)&=&  \pi\,,
\quad
\widetilde{F}'(y_{0}^2) = \alpha\,,
\\[2mm]
\label{eq:BCS-NUM-sigmatilde-nutilde}
\widetilde{\sigma}(y_{0}^2)&=& \beta\,,
\quad
\widetilde{\mu}(y_{0}^2)= \gamma\,,
\end{eqnarray}
\esubeqs
for certain initial values $\alpha,\,\beta,\,\gamma \in \mathbb{R}$.
With these boundary conditions, the ODEs \eqref{eq:final-ODEs} are solved over
$w\in [y_0^2,\,w_\text{max}]$ for sufficiently large values of $w_\text{max}$.
Next, the numbers $\alpha$ and $\beta$ in \eqref{eq:BCS-NUM}
are varied in order to obtain
vanishing $\widetilde{F}(w)$ and $\widetilde{F}^\prime(w)$ at $w=w_\text{max}$.
The obtained function $\widetilde{\mu}(w)$ can then be rescaled
in order to obtain at $w=w_\text{max}$ the Schwarzschild-type values  
\bsubeqs\label{eq:Schwarzschild-behavior}
\begin{eqnarray}
\label{eq:Schwarzschild-behavior-sigma}  
\widetilde{\sigma}(w_\text{max})&\equiv&
\left(\sqrt{1-l_\text{max}/\sqrt{w_\text{max}}}\,\right)^{-1}\,,
\\[2mm]
\label{eq:Schwarzschild-behavior-mu}
\widetilde{\mu}(w_\text{max})&=&
\sqrt{1-l_\text{max}/\sqrt{w_\text{max}}}\,,
\end{eqnarray}
\esubeqs
with a dimensionless length parameter $l_\text{max}$
set by the value of $\widetilde{\sigma}(w_\text{max})$.
[An alternative numerical procedure would be to directly
use the boundary conditions \eqref{eq:BCS-MATH},
but then we would have to solve a so-called
``two-point boundary value problem'' instead of
the ``initial value problem'' from \eqref{eq:BCS-NUM}.]

Numerical results for the dimensionless gravitational coupling constant
$\widetilde{\eta}=1/20$ and the dimensionless defect size $y_0=1/\sqrt{2}$
are given in Figs.~\ref{fig:functions-bmid}--\ref{fig:core-bmid}.
Several comments are in order.
First, the results from the bottom row panels in
Fig.~\ref{fig:functions-bmid} make clear that the solution
is localized.
Second, the middle and right panels of the top row
in Fig.~\ref{fig:asymptotics-bmid} show the Schwarzschild
behavior \eqref{eq:Schwarzschild-behavior} for $w\gtrsim 10$.
Third, it appears that the energy density  
$-T^{0}_{\;\;0} = -E^{0}_{\;\;0}/\widetilde{\eta}$
from the bottom right panel of Fig.~\ref{fig:asymptotics-bmid}
behaves asymptotically as $1/w^2 \sim 1/|y|^4$,
which results in a finite energy integral.
Fourth, the physical quantities shown in the
bottom row of Fig.~\ref{fig:core-bmid} are nonsingular 
at the defect core ($y=0$),  
in accordance with the analytic expressions \eqref{eq:R-K-E00-expressions}.

\begin{table*}[t]
\vspace*{0mm}
\begin{center}
\caption{\label{tab:masses}%
Numerical results for the
dimensionless Schwarzschild mass $\widehat{m}\equiv 4\pi l/\widetilde{\eta}$
in the theory with dimensionless gravitational
coupling constant $\widetilde{\eta}=1/20$.
The dimensionless length parameter $l$ is obtained from the
numerical solution by use of \eqref{eq:Schwarzschild-behavior-mu}.
Also shown are the values of the
flat-spacetime energy integral $E^\text{flat}$
relative to its Bogomolnyi  value $12\, \sqrt{2}\, \pi^2\, f/e$.
This integral $E^\text{flat}$
depends only on the Skyrme function $\widetilde{F}(w)$ and has
both metric functions
$\widetilde{\sigma}(w)$ and $\widetilde{\mu}(w)$ set to unity;
see Appendix~\ref{app:Reduced-expressions}.
As such, $E^\text{flat}[\widetilde{F}]$ can be used as a diagnostic of
the Skyrme function $\widetilde{F}(w)$.
\vspace*{5mm}}
\renewcommand{\tabcolsep}{1.1pc}    %% enlarge column spacing
\renewcommand{\arraystretch}{1.1}   %% enlarge line spacing
\begin{tabular}{l|ccc}
\hline\hline
$y_0$
& $l/2$
& $\widehat{m}$
& $E^\text{flat}/E^\text{flat}_\text{Bogomolnyi}$
\\
\hline
$1/(2\sqrt{2})$
& $-$
& $-$
& $-$
\\
$1/2$
& $0.313$
& $78.7 $
& $2.41 $
\\
$1/\sqrt{2}$
& $0.426$
& $107 $
& $2.27 $
\\
$1$
& $0.669 $
& $168 $
& $1.79 $
\\
$\sqrt{2}$
& $1.21 $
& $304 $
& $1.70 $
\\
$2$
& $1.58 $
& $398 $
& $2.51 $
\\
\hline\hline
\end{tabular}
\end{center}
%%\vspace*{1000mm}%%tmp
\end{table*}

Similar results have been obtained for other values
of the dimensionless defect size $y_0$, provided they are not too small:
$y_0 > y_{0,\,\text{crit}}$ with $y_{0,\,\text{crit}} \sim 1/(2\sqrt{2})$
for coupling constant
$\widetilde{\eta}=1/20$.\footnote{\label{ftn:y0crit}The
existence of this critical defect size $y_{0,\,\text{crit}}$
remains to be confirmed.}
Table~\ref{tab:masses} gives the corresponding values for the
dimensionless Schwarzschild mass $\widehat{m}$.
Qualitatively, these $\widehat{m}$ values agree with those
of Fig.~8.7b in Ref.~\cite{Schwarz2010}, but not quantitatively.
This quantitative difference most likely traces back to the fact that
the fields of Ref.~\cite{Schwarz2010}
do not solve the Einstein equations at the defect core
($W=b^2$ in our notation).  This then results in a different
behavior of the fields further out, $W>b^2$, making for
different numerical values of $\widehat{m}$.

The present paper presents only exploratory numerical results,
with the sole purpose of establishing the existence of at least one type of
Skyrmion-like spacetime-defect solution.
For now, two branches of solutions have been found,
distinguished by the value of the slope of the Skyrme function
at the defect core, $\widetilde{F}^\prime(y_0^2)$.
The branch with smaller $|\widetilde{F}^\prime(y_0^2)|$
has larger Schwarzschild mass $\widehat{m}$
(the numerical results for this branch can be found in a previous version of
this paper~\cite{Klinkhamer2014-v2}).  
The branch with larger $|\widetilde{F}^\prime(y_0^2)|$
has smaller Schwarzschild
mass\footnote{\label{ftn:explanation}A possible
heuristic explanation of this behavior relies on
embedding the $\text{deg}[\Omega]=1$ defect solution
[having $W \geq b^2 >0$ and $\widetilde{F}(b^2)=\pi$]
in the $\text{deg}[\Omega]=2$ solution without defect
[having $W \geq 0$ and $\widetilde{F}(0)=2\pi$].
A larger value of $-\widetilde{F}^\prime>0$
at the point where $\widetilde{F}=\pi$
then results in a smaller value of $b$ with correspondingly smaller
Schwarzschild mass value $\widehat{M}$ (cf. Table~\ref{tab:masses}).}
and the numerical results of this branch have been given in
the present paper. More numerical work is clearly needed.

%%\newpage%%tmp
\section{Discussion}
\label{sec:Discussion}

In this article, we have succeeded in
constructing a nonsingular localized finite-energy solution of
the standard Einstein equations with
nontrivial topology on small length scales and
flat spacetime asymptotically.
The particular topology is given by \eqref{eq:M4-M3-topology}
and the matter content by a Skyrme model with
action \eqref{eq:action}.
The three crucial inputs for obtaining the solution are
first, a proper set of coordinates (Sec.~\ref{subsec:Manifold});
second, appropriate \textit{Ans\"{a}tze} (Sec.~\ref{subsec:Ansaetze})
and, third, special attention to the behavior of the
fields at the defect core for the numerical
solution (Sec.~\ref{sec:Numerical-solution}).

This new type of Skyrmion solution is rather interesting:
it combines the nontrivial topology of spacetime with the
nontrivial topology of field-configuration space.
The interplay of spacetime and internal space
involves the standard hedgehog behavior
(well known from the magnetic-monopole, sphaleron,
and standard Skyrmion solutions) and
the nontrivial topology of the underlying space manifold
allows the internal $SO(3)$ space to be covered only once 
(see the discussion in Footnote~\ref{ftn:explanation}).  

The Skyrmion defect solution of this paper resembles in certain
aspects the defects discussed in brane-world
scenarios~\cite{DvaliKoganShifman2000,CembranosDobadoMaroto2002}.
But, unlike the brane-Skyrmion of Ref.~\cite{CembranosDobadoMaroto2002},
our Skyrmion defect solution does not
neglect any part of the gravitational interactions, it is a
complete solution of the theory considered.

It remains to be proved that the solution obtained in the
present article is stable.
The scalar fields by themselves would
be stable because of the topological charge \eqref{eq:deg-Omega},
but, in principle, there could be still more branches of solutions
with even lower values of the Schwarzschild mass.
A rigorous proof for the stability of the self-gravitating
solution would be most welcome.
(Equally welcome would be a rigorous proof for the
existence of the self-gravitating solution, as we have
relied on a numerical analysis of the reduced field equations.)

In Refs.~\cite{Klinkhamer2013-MPLA,Klinkhamer2013-APPB},
we have given a heuristic discussion on how the corresponding nonsingular
black-hole solutions could appear in physical situations of
spherical gravitational collapse in an essentially flat spacetime
with trivial topology. The scenario is that, when the central matter
density becomes large enough, there is a quantum
jump~\cite{Wheeler1957,Wheeler1968}
from the trivial $\mathbb{R}^4$ topology to the
nonsimply-connected $M_4$ topology of the nonsingular solution,
with a noncontractible-loop length scale $b$
of the order of the quantum-gravity length scale
$L_\text{Planck}$ $\equiv$ $(\hbar\, G_N/c^3)^{1/2}$.  
(Perhaps topology change is not needed if the
nontrivial small-scale topology is already
present as a remnant from a quantum spacetime
foam~\cite{Klinkhamer2013-review}.)

If a similar discussion applies to the
case of the Skyrmion spacetime defect found in this paper,
it can be conjectured that defects with the smallest possible
value of the mass $\widehat{M}$
would have the largest probability of occurring.
For the solutions of Table~\ref{tab:masses}
and setting $\hbar=c=1$, this would imply  
a preferred value of the defect length scale of
approximately $b\equiv y_0/(e f) \sim 0.5/(e f)$
for a defect mass of approximately
$\widehat{M}\equiv \widehat{m}\,f/e \sim 80\,f/e$, with
scalar-field energy scale
$f=\sqrt{1/20}\,E_{p}$ in terms of the reduced Planck
energy $E_{p} \equiv 1/\sqrt{8\pi\, G_N}
\approx  2.4 \times 10^{18}\,\text{GeV}$.

At this moment, we should mention that
the metric \eqref{eq:metric-Ansatz-W-definition}
of the nonsingular defect solution does have a
blemish~\cite{Klinkhamer2013-MPLA,Klinkhamer2013-review}:
at $W=b^2$, it cannot be brought to a patch of Minkowski spacetime
by a genuine diffeomorphism
(a $C^\infty$ function) but only by a $C^1$ function.
This may very well be the price to pay for having
a nonsingular solution. The ultimate quantum theory
of spacetime and gravity must                                determine which role
(if any) these nonsingular defect-type classical solutions play
for the origin of classical spacetime in the very early universe. 

%%\newpage%%tmp
\section*{\hspace*{-5mm}ACKNOWLEDGMENTS}
\vspace*{-0mm}\noindent
It is a pleasure to thank C. Rahmede for discussions
and A. de la Cruz-Dombriz for mentioning brane-Skyrmions.

\begin{appendix}
\section{Alternative Ans\"{a}tze}
\label{app:Alternative-Ansaetze}

In this appendix, we present alternative \textit{Ans\"{a}tze}, possibly
relevant for nonsingular Skyrmion black-hole solutions
with topology \eqref{eq:M4-M3-topology} and defect length scale $b$.

The new form of the metric \textit{Ansatz} is related to the
one of \eqref{eq:metric-Ansatz-W-definition}
by the introduction of a new time
coordinate, $\widehat{T}=\widehat{T}(T,\,Y_2)$.
This new coordinate follows from considering freely-moving observers 
falling in along  radial geodesics and
writing their covariant four-velocities as gradients of a
new time function $\widehat{T}$.
As such, the procedure is analogous to that of
changing the standard coordinates of the Schwarzschild
metric to the Painlev\'{e}--Gullstrand
coordinates~\cite{Painleve1921,Gullstrand1922,MartelPoisson2000},
generalized to other static spherically symmetric spacetimes
(cf. Sec. IV of Ref.~\cite{MartelPoisson2000}).

In this way, we arrive at the following \textit{Ansatz}
for the line element:
\bsubeqs\label{eq:metric-alternative-Ansatz-W-definition}  
\beqa\label{eq:metric-alternative-Ansatz}
ds^2\;\Big|_\text{chart-2} &=&
- d\widehat{T}^2
+ \Bigg[\widetilde{\chi}(W)\, \frac{Y_{2}}{\sqrt{W}}\;\,dY_2
+ \widetilde{\xi}(W)\, d\widehat{T}\,\Bigg]^2
\nonumber\\
&&
 +W \Big[(dZ_2)^2+\sin^2 Z_2\, (dX_2)^2 \Big]\,,
%%\\[2mm]
\eeqa
\beqa
W\;\Big|_\text{chart-2} &\equiv& b^2+(Y_{2})^2\,,
\eeqa
\esubeqs
with functions $\widetilde{\chi}(W)$ and $\widetilde{\xi}(W)$.
Similar \textit{Ans\"{a}tze} hold  for the chart-1 and chart-3 domains;
see Appendix C of Ref.~\cite{Klinkhamer2013-review}.
The \textit{Ansatz} for the scalar $SO(3)$ field takes the same
form as the one presented in Sec.~\ref{subsec:Ansaetze}.

The  boundary conditions for the Skyrme function $\widetilde{F}(W)$
are given by \eqref{eq:hedgehog-Ansatz-bcs}
and those for the new metric functions
$\widetilde{\chi}(W)$ and $\widetilde{\xi}(W)$  by
\beq
\widetilde{\chi}(\infty)=1\,,\quad
\widetilde{\xi}(\infty)=0\,,
\eeq
which correspond to Minkowski spacetime asymptotically.

%%\newpage%%tmp
\section{Reduced expressions}
\label{app:Reduced-expressions}

In this appendix, we give the reduced expressions
for certain curvature tensors. In addition, we give the
expression for the flat-spacetime energy integral used in Table~\ref{tab:masses}.

Specifically,
consider the  Ricci scalar $R \equiv g^{\mu\nu}\,R_{\mu\nu}$,
the Kretschmann scalar
$K$ $\equiv$ $R_{\mu\nu\rho\sigma}\,R^{\mu\nu\rho\sigma}$,
and the 00 component of the Einstein tensor
$E^{\mu}_{\;\;\nu}$ $\equiv$ $R^{\mu}_{\;\;\nu} -(1/2)\,R\,\delta^{\mu}_{\;\;\nu}$.
Then, the \textit{Ansatz} from Sec.~\ref{subsec:Ansaetze} produces the
following expressions:
\bsubeqs\label{eq:R-K-E00-expressions}
\begin{eqnarray}
\label{eq:R-expression}
R(w)&=&
2 \,(e\,f)^2\;
\Big(
\widetilde{\sigma}(w)^3 -\widetilde{\sigma}(w)
+ 4 \,w\, \widetilde{\sigma}'(w)
- 6 \,w\, \widetilde{\sigma}(w)\, \widetilde{\mu}'(w)/\widetilde{\mu}(w)
\nonumber\\&&
+ 4 \,w^2 \,\widetilde{\sigma}'(w)\,\widetilde{\mu}'(w)/\widetilde{\mu}(w)
- 4 \,w^2\, \widetilde{\sigma}(w)\, \widetilde{\mu}''(w)/\widetilde{\mu}(w)
\Big)\Big/
\Big(w \, \widetilde{\sigma}(w)^3\Big)\,,
%\\[2mm]
\end{eqnarray}\begin{eqnarray}
\label{eq:K-expression}
K(w)&=&
4 \,(e\,f)^4\;
\Big(
\widetilde{\sigma}(w)^2 - 2 \,\widetilde{\sigma}(w)^4
+ \widetilde{\sigma}(w)^6 + 8 \,w^2\,  \widetilde{\sigma}'(w)^2
\nonumber\\&&
+ 12\, w^2\, \widetilde{\sigma}(w)^2\, \widetilde{\mu}'(w)^2/\widetilde{\mu}(w)^2
- 16\, w^3\, \widetilde{\sigma}(w)
\, \widetilde{\sigma}'(w)\,\widetilde{\mu}'(w)^2/\widetilde{\mu}(w)^2
\nonumber\\&&
+ 16\, w^4\,  \widetilde{\sigma}'(w)^2\, \widetilde{\mu}'(w)^2/\widetilde{\mu}(w)^2
\nonumber\\&&
+ 16\, w^3\, \widetilde{\sigma}(w)^2\, \widetilde{\mu}''(w)\,
  \widetilde{\mu}'(w)/\widetilde{\mu}(w)^2
- 32\, w^4\, \widetilde{\sigma}(w)\,
  \widetilde{\sigma}'(w)\,\widetilde{\mu}''(w)\,\widetilde{\mu}'(w)/\widetilde{\mu}(w)^2
  \nonumber\\&&
+ 16\, w^4\, \widetilde{\sigma}(w)^2\, \widetilde{\mu}''(w)^2/\widetilde{\mu}(w)^2
\Big)\Big/
\Big(w^2\, \widetilde{\sigma}(w)^6\Big)\,,
%\\[2mm]
\end{eqnarray}\begin{eqnarray}
\label{eq:E00-expression}
E^{0}_{\;\;0}(w)&=&
(e\,f)^2\;
\Big(\widetilde{\sigma}(w) - \widetilde{\sigma}(w)^3
- 4\, w\, \widetilde{\sigma}'(w)\Big)\Big/
\Big(w\, \widetilde{\sigma}(w)^3\Big)\,.
\end{eqnarray}
\esubeqs
We have set $(e\,f)=1$ for the results shown in
Figs.~\ref{fig:functions-bmid}--\ref{fig:core-bmid}.

Evaluating the negative of the matter Lagrange density 
from \eqref{eq:action} with the \textit{Ansatz} fields and
setting the \textit{Ansatz} metric functions
$\widetilde{\sigma}(w)$ and $\widetilde{\mu}(w)$ to unity
gives the flat-spacetime energy integral $E^\text{flat}$
used in Table~\ref{tab:masses}:
\begin{eqnarray}
\label{eq:Eflat-expression}
E^\text{flat}
&=&
4\pi \;(f/e)\;\int_{y_{0}^2}^{\infty}\;dw\;w^{-3/2}\;
\Big[ \Big(1  +\cos[\widetilde{F}(w)]+ 2 w\Big)\, \sin^2[\widetilde{F}(w)/2]
\nonumber\\&&
+ w^2\, \Big(2  - 2 \cos[\widetilde{F}(w)]+ w\Big)\, \widetilde{F}'(w)^2\Big]\,,
\end{eqnarray}
which agrees with previous
results~\cite{Skyrme1961,MantonSutcliffe2004,Schwarz2010}.
Part of the integral is proportional to the absolute value of the
topological degree $N$, with an integral of nonnegative terms remaining.   
This results in the following Bogomolnyi-type inequality~\cite{Schwarz2010}:
\beq\label{eq:Eflat-lower-bound}
E^\text{flat}
\;\geq\;
12\, \sqrt{2}\, \pi^2\, |N|\,f/e
\;\equiv\;
E^\text{flat}_\text{Bogomolnyi} \,.
\eeq
The inequality \eqref{eq:Eflat-lower-bound} for $N=1$
is not saturated by the self-gravitating Skyrmion defect solution,
according to the numerical results from Table~\ref{tab:masses}.
The same holds for the nongravitating flat-spacetime
$SO(3)$ Skyrmion defect solution~\cite{Schwarz2010} and 
the standard Minkowski-spacetime $SU(2)$ Skyrmion~\cite{MantonSutcliffe2004}.

\end{appendix}

%%\newpage%%tmp

\end{document}